\documentclass[12pt]{article}

\usepackage[breaklinks]{hyperref}  
\usepackage{graphicx}
\usepackage[active]{srcltx} 
\usepackage{sariel,wide}
\usepackage{amstext}
\usepackage{amsmath}
\usepackage{picins}

\DefineNamedColor{named}{OliveGreen}    {cmyk}{0.64,0,0.95,0.40}

\newcommand{\PntSet}{P}
\newcommand{\LinesX}[1]{\EuScript{L}\pth{#1}}

\newcommand{\pnt}{\mathsf{p}}
\newcommand{\pntA}{\mathsf{q}}
\newcommand{\pntB}{\mathsf{r}}

\newcommand{\lineA}{\ell}

\newcommand{\pairsX}[1]{\mathrm{E}\pth[]{#1}} 

\newcommand{\Term}[1]{\textsf{#1}}
\newcommand{\LP}{\Term{L{}P}\xspace}
\newcommand{\VC}{\Term{V{C}}\xspace}
\newcommand{\Family}{\EuScript{F}}

\newcommand{\ArrX}[1]{\mathop{\mathrm{\EuScript{A}}}\pth{#1}}

\newcommand{\Dim}{\tau}

\newcommand{\dCr}[2]{d_{#1}(#2)}
\newcommand{\diskCr}[2]{\mathsf{D}_{#1}(#2)}
\newcommand{\LineSet}{L}

\providecommand{\Holder}{H\"older\xspace}
\providecommand{\Bronnimann}{Br{\"o}nnimann\xspace}
\newcommand{\I}{\mathcal{I}}

\begin{document}

\title{Approximating Spanning Trees with Low Crossing Number}

\author{Sariel Har-Peled\SarielThanks{}}

\date{\today}

\maketitle

\begin{abstract}
    We present a linear programming based algorithm for computing a
    spanning tree $T$ of a set $\PntSet$ of $n$ points in $\Re^d$,
    such that its crossing number is $O(\min( t \log n, n^{1-1/d}) )$,
    where $t$ the minimum crossing number of any spanning tree of
    $\PntSet$. This is the first guaranteed approximation algorithm
    for this problem.  We provide a similar approximation algorithm
    for the more general settings of building a spanning tree for a
    set system with bounded \VC dimension.

    Our approach is an alternative to the reweighting technique
    previously used in computing such spanning trees.
\end{abstract}


\section{Introduction}

The reweighting technique is a powerful tool in computer science
\cite{ahk-mwmma-06}. In Computational Geometry, it was introduced by
Chazelle and Welzl \cite{cw-qorss-89} who used it to compute spanning
paths with low crossing number in set systems with bounded \VC
dimension. Welzl \cite{w-stlcn-92} provided a tighter analysis for the
case of spanning tree of points in $\Re^d$.  \Matousek \cite{m-ept-92}
used the reweighting technique to provide a powerful partition theorem
that proved to be very useful in building range searching
data-structures \cite{ae-rsir-98}.  Also, Clarkson \cite{c-apca-93}
provided an algorithm for polytope approximation that used the
reweighting technique.  \Bronnimann and Goodrich \cite{bg-aoscf-95}
realized that Clarkson's algorithm implies a general method for
solving hitting set and set cover problems in geometric settings.

Interestingly, Long \cite{l-upsda-01} had observed that set cover
problems in geometric settings can be solved by using \LP and taking a
random sample (guided by the \LP solution) that is an $\eps$-net (a
similar observation was later made by \cite{ers-hsvcs-05}). In fact,
such packing/covering \LP{}s can be solved efficiently via reweighting
\cite{pst-faafp-91}. Thus, one can interpret the reweighting algorithm
for solving the geometric set cover problem as directly solving the
associated \LP.

\parpic[r]{\includegraphics{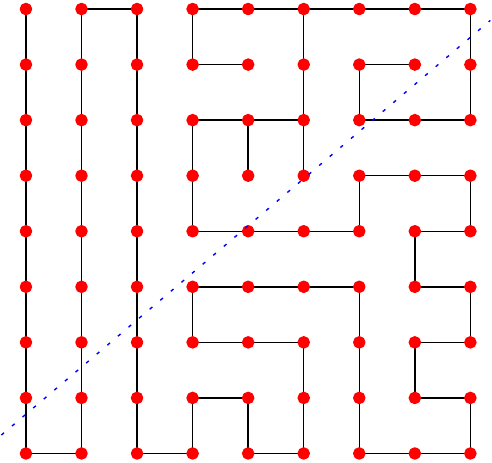}}

The result of Welzl \cite{w-stlcn-92}, mentioned above, is quite
intriguing. It shows that for a set $\PntSet$ of $n$ points in the
plane (resp., in $\Re^d$) one can find a spanning tree of the points,
such that any line (resp., hyperplane) crosses at most $O(\sqrt{n})$
(resp., $O(n^{1-1/d})$) edges (i.e., segments) of the spanning tree.
To appreciate this result, consider the point set formed by the grid
$\sqrt{n} \times \sqrt{n}$. It is trivial in this case to come up with
a spanning tree with a crossing number $O( \sqrt{n})$ -- any spanning
tree of the grid points using only edges of the grid has this
property, see figure on the right. Surprisingly, the result of Welzl
\cite{w-stlcn-92} implies that any point set behaves like a grid point
set as far as the crossing number of the optimal spanning tree.

\bigskip In this work, we establish a connection between computing
spanning trees with low crossing number and \LP{}s (i.e., linear
programs); that is, we show that spanning trees with low crossing
number can be computed using \LP rounding.

\medskip
\noindent
\textbf{Approximate spanning tree with the lowest crossing number.}
Given a set system $\I = (\PntSet, \Family)$ of finite \VC dimension
$\Dim$, we show how to compute, in polynomial time%
\footnote{We make the standard assumption that solving a \LP of
   polynomial size takes polynomial time.}%
, a spanning tree of $\PntSet$ with crossing number $O(t \log n)$
(assuming $t =\Omega( \log n)$), where $t$ is the minimal crossing
number of any spanning tree of $\PntSet$.  This is done by recursively
solving a \LP relaxation and rounding it.  See \secref{approximation}
for details.

Naturally, this algorithm also applies to the Euclidean case.
Specially, given a set $\PntSet$ of $n$ points in $\Re^d$ one can
compute, in polynomial time, a spanning tree $T$ such that every
hyperplane crosses at most $O(t \log n)$ edges of $T$, where $t$ is
the minimum crossing number of any spanning tree of $\PntSet$.  

Surprisingly, this is the first guaranteed approximation algorithm
known for this problem. In particular, achieving such an approximation
is mentioned as open in the Open Problems Project (see
\url{http://maven.smith.edu/~orourke/T%
   O%
   P%
   P/P20.html#Problem.20}).

\medskip
\noindent
\textbf{Spanning trees in $\Re^d$ with $O(n^{1-1/d})$ crossings.}  We
also modify the analysis of our algorithm (but not the algorithm
itself) so that it yields worst case bound on the crossing number.
Specifically, we get a polynomial time algorithm that, for a given set
$\PntSet$ of $n$ points in $\Re^d$, computes a spanning tree $T$ of
$\PntSet$ such that any hyperplane in $\Re^d$ crosses at most
$O(n^{1-1/d})$ edges of $T$.  Our proof of the correctness of the
algorithm is self contained (except for a relatively easy lemma, see
\lemref{crossing:disk}), and uses \LP duality. We believe the new
proof provides a new insight into why such trees exist. In particular,
Chazelle and Welzl \cite{cw-qorss-89} and Welzl \cite{w-stlcn-92}
proofs of the existence of such spanning trees are simple but somewhat
``mysterious'' (at least for the author, but other people might not
see the mystery).

Here is a sketch of the resulting argument why such trees exist: In
the plane, it is sufficient to find spanning forest that span at least
$\Omega(n)$ vertices of $\PntSet$ (in connected components that are
not singletons) and has crossing number $t$.  One can find such a
(fractional) spanning graph by doing \LP relaxation. The dual \LP then
asks (intuitively) to separate the given $n$ points into singletons by
a set of lines of minimum cardinality (i.e., for any $\pnt, \pntA \in
\PntSet$, there exists a selected line that crosses the segment
$\pnt\pntA$). It is not hard to show that any such set of lines need
to be of size $\Omega(\sqrt{n})$. A somewhat more involved argument
(since we are dealing with a fractional solution of an \LP that has
some other constraints) implies that the dual \LP is feasible for $t =
\sqrt{n}$ and its optimal solution is bounded from below. It follows
that the primal \LP is feasible. Now
, solving the primal \LP and using
a straightforward rounding implies that one can compute the required
spanning graph. Applying this recursively by selecting a vertex from
each connected component, and overlaying the resulting spanning graphs
together results in a connected graph of $\PntSet$ with crossing
number $O(\sqrt{n})$.  See \secref{planar} for details.

Interestingly, while the above algorithm works for any point set in
$\Re^d$, one can do slightly better in the planar case, and get a
deterministic rounding scheme, see \secref{deterministic} for details.

\medskip

\paragraph{Previous work.} Fekete \etal \cite{flm-msnmt-08} suggested
using \LP relaxation to compute a spanning tree with low crossing
number. Their \LP is considerably more elaborate than ours
(considering all cuts), and their iterated rounding scheme seems to
perform quite well in practice (although they are unable to provide a
theoretical guarantee on the performance). Furthermore, they prove
that computing the spanning tree with minimal crossing number is
\NPHard.

\paragraph{Organization.}
In \secref{approximation} we show the $O(t \log n)$ approximation
algorithm for spanning trees with low crossing number, for the general
case of a set system with low \VC dimension.  In \secref{planar}, we
specialize this algorithm for the case of points and hyperplanes in
$\Re^d$.  We discuss our results and some related open problems in
\secref{conclusions}.

\section{Approximating the spanning tree with optimal crossing number}
\seclab{approximation}

Consider a set system $\I = (\PntSet, \Family)$ of finite \VC
dimension $\Dim$. For more details on spaces with bounded \VC
dimension see \cite{pa-cg-95}. For our purposes, it is sufficient that
$\Family$ is a set of subsets of $\PntSet$ of cardinality bounded by
$\cardin{\PntSet}^\Dim$, and this holds for any set system induced by
a subset of $\PntSet$.

For two distinct points $\pnt, \pntA \in \PntSet$, we will refer to
the set $\brc{ \pnt, \pntA}$ as an \emphi{edge}, denoted by $\pnt
\pntA$.  An edge $\pnt \pntA$ \emphi{crosses} a set $S \in \Family$ if
$\cardin{ \brc{\pnt, \pntA} \cap S} = 1$.

The \emphi{crossing number} of a set of edges $F$ of $\PntSet$ is the
maximum number of edges of $F$ crossed by any set of $\Family$.

\begin{example}
    As a concrete example of such a set system, consider a set $\PntSet$
    of $n$ points in the plane, and let 
    \[
    \Family = \brc{ \PntSet \cap h^+ \sep{ h^+ \text{ is a halfplane}}}.
    \]
    The set system $\I = (\PntSet, \Family)$ in this case has \VC
    dimension $3$, and a spanning tree $T$ of $\PntSet$ with crossing
    number $t$, is a spanning tree of $\PntSet$, drawn in the plane by
    straight segments, such that every line (i.e., the boundary of a
    halfplane) intersects at most $t$ edges of $T$.
\end{example}

\begin{lemma}
    Assume there exists a spanning tree $T$ for $\PntSet$ with
    crossing number $t$. Then, for any subset $X \subseteq \PntSet$
    there exists a spanning tree with crossing number at most $2t$.

    \lemlab{inherent}
\end{lemma}

\begin{proof}
    Convert the spanning tree $T$ of $\PntSet$, with crossing number
    $t$, into a closed cycle $C$ visiting the points of $\PntSet$, by
    doing an Euler tour of $T$, using each edge of $T$ twice. The new
    cycle $C$ has crossing number $2t$.  Next, shortcut the cycle $C$
    such that it uses only elements of $X$, by replacing each subpath
    $\pi_{xy}$ (that uses inner vertices that are not in $X$)
    connecting $x,y \in X$ by the edge $xy$.

    This results in a cycle that visits only the vertices of $X$ and
    has crossing number $\leq 2t$, as such a shortcutting can only
    decrease the number of edges crossing a set $S \in
    \Family$. Indeed, consider a subpath $x_1 x_2, x_2x_3, \ldots,
    x_{m-1}x_m$, and observe that if $x_1x_m$ crosses a set $S$, then
    one of the edges in the path must also cross this set. Thus,
    replacing a subpath (of a cycle) by an edge reduces the crossing
    number of the cycle.
\end{proof}

\bigskip

Let $\pairsX{\PntSet}$ denote the set of all edges of $\PntSet$, and
consider the following \LP (parameterized by $t$).
\begin{align}
    \gamma(\PntSet,t) = \max & \sum_{\pnt\pntA \in \pairsX{\PntSet}}
    y_{\pnt \pntA} & \eqlab{l:p:general}  \\
    \nonumber s.t. \quad\quad & \sum_{\substack{\pnt \pntA \in
          \pairsX{\PntSet},
          \cardin{\pnt\pntA \cap S} = 1}} y_{\pnt\pntA} \leq t &
    \forall S \in \Family \\
    \nonumber
    & \sum_{\substack{ \pntA \in \PntSet,
          \pntA \ne \pnt}}
    y_{\pnt\pntA} \geq 1 & \forall \pnt \in
    \PntSet \tag{*} \\
    \nonumber
    & y_{\pnt\pntA} \geq 0 & \forall \pnt\pntA \in \pairsX{\PntSet}.
\end{align}

Intuitively, this \LP tries to pick as many edges as possible (i.e.,
$y_{\pnt \pntA} =1$ indicates that we pick the edge $\pnt \pntA$),
such that (i) no set is being crossed more than $t$ times, and (ii)
every point of $\PntSet$ participates in at least one edge that is
being picked.

\begin{remark}
    The above \LP might not be feasible, and in such a case
    $\gamma(\PntSet, t)$ is not defined. Naturally, one can modify
    this \LP to first compute the minimal value $t$ for which it is
    feasible, and then solve the original \LP with this value of $t$.

    In particular, in this case, we can choose almost any arbitrary
    target function to optimize the \LP for. We had chosen this one
    since the dual form is convenient to work with, see
    \secref{planar}.
\end{remark}

\begin{lemma}
    Consider a set system $\I = (\PntSet, \Family)$ with bounded \VC
    dimension, where $n = \cardin{\PntSet}$, and let $t$ be a
    parameter such that $\gamma(\PntSet,t)$ is feasible. Then, one can
    compute (in polynomial time) a set of edges $F$, such that
    $\Ex{\cardin{F}} = \gamma(\PntSet,t)$, and the number of connected
    components of the graph $(\PntSet, F)$ is (in expectation) at most
    $(9/10)n$.  The crossing number of $F$ is $O( t + \log n / \log
    \log n)$ with high probability.

    \lemlab{spanning}
\end{lemma}
\begin{proof}
    We solve the \LP \EqrefQ{l:p:general} and compute $\gamma(\PntSet,
    t)$. Next, for every $\pnt \pntA \in \pairsX{\PntSet}$, if
    $y_{\pnt \pntA} \geq 1$ then we add $\pnt \pntA$ to
    $F$. Otherwise, if $y_{\pnt \pntA} < 1$ then we pick the edge
    $\pnt \pntA$ into $F$ with probability $y_{\pnt\pntA}$.
    
    For a set $S \in \Family$, let $X_S$ be its crossing number in
    $F$.  We have that
    \[
    \mu = \Ex{X_S} \leq \sum_{\substack{\pnt \pntA \in \pairsX{\PntSet},
          \\ \cardin{\pnt\pntA \cap S} = 1}} y_{\pnt\pntA} \leq t.
    \]
    For a constant $c > 0$ sufficiently large, let $\displaystyle
    \delta = c + c \pth[]{ \log n}/\pth{t \log \frac{\log n}{t} }$,
    and observe that $t \delta \log \delta = \Omega( \log n)$. As
    such, by the Chernoff inequality, we have that
    \begin{align*}
        \Prob{X_S > (1+\delta)t }
        &\leq \Prob{X_S > \pth{1+\frac{t\delta}{\mu}}\mu } \leq
        \pth{\frac{\exp(t\delta/\mu)}{(1+t\delta/\mu)^{1+t\delta/\mu}}}^\mu
        \\
        & = 
        \exp \pth{t \delta - (\mu + t \delta) \ln \pth{ 1+ \frac{
                 t \delta}{\mu}} }
        \leq 
        \exp \pth{t \delta -  t \delta \log \delta }
        < \frac{1}{n^{O(1)}}.
    \end{align*}
    Since $\I$ has \VC dimension $\Dim$, the number of sets one has to
    consider (i.e., the size of $\Family$) is $O \pth{ n^{\Dim}}$
    \cite{pa-cg-95}, which implies, by the above, that the crossing
    number of $F$ is bounded by $(1+\delta)t = O( t + \log n / \log
    \log n)$, with high probability.

    As for the number of connected components in the graph $G =
    (\PntSet, F)$, observe that a point $\pnt$ is not adjacent to any
    edge of $F$ with probability
    \[    
    \prod_{\substack{ \pntA \in \PntSet,\\ \pntA \ne \pnt}}
    (1 - y_{\pnt\pntA}) 
    \leq \exp \pth{ -\sum_{{ \pntA \in \PntSet, \pntA \ne \pnt}}
        y_{\pnt\pntA}} \leq \frac{1}{e},
    \]
    by the inequality (*) in \LP \EqrefQ{l:p:general}. Let $Y$ be the
    number of points of $\PntSet$ that are singletons in the graph
    $(\PntSet,F)$. By the above, we have that $\Ex{Y} \leq n/e$. As
    such, the expected number of connected components in $(\PntSet,F)$
    is
    \[
    \leq \Ex{\frac{n-Y}{2} + Y} \leq \frac{n}{2} + \frac{n}{e} \leq
    \frac{9}{10} n.
    \]
    \aftermathA
\end{proof}

\begin{theorem}
    Consider a set system $\I = (\PntSet, \Family)$ with bounded \VC
    dimension, where $n = \cardin{\PntSet}$.  Let $t$ be the minimum
    crossing number of any spanning tree of $\I$.  Then, one can
    compute, in polynomial time, a spanning tree $T$ of $\PntSet$ with
    a crossing number $O(t \log n + \log^2 n / \log \log n )$.

    \thmlab{low:crossing}
\end{theorem}
\begin{proof}
    We set $\PntSet_1 = \PntSet$. In the $i$th iteration, we compute
    the minimal $t_i$ for which $\gamma(\PntSet_i,t_i)$ is
    feasible. Next, we compute a set of edges $F_i$ over $\PntSet_i$,
    using \lemref{spanning}. If the number of connected components of
    $(\PntSet_i, F_i)$ is larger than $(19/20) \cardin{\PntSet_i}$, we
    repeat this iteration (we have constant probability to succeed by
    Markov's inequality). Next, from each connected component of
    $\PntSet_i$, we pick one point into $\PntSet_{i+1}$. We repeat
    this algorithm till we remain with a single point. This algorithm
    performs $m = O( \log n)$ iteration.  Now, the union $F = \cup_i
    F_i$ forms a spanning graph of $\PntSet$, and we return any
    spanning tree $T$ of $(\PntSet,F)$.

    The crossing number of $T$ is bounded by the total crossing
    numbers of the graphs $G_1 = (\PntSet_1, F_1), \ldots, G_m =
    (\PntSet_m, F_m)$. Now, the graph $G_i$ has crossing number $O(
    t_i + \log n / \log \log n)$ by \lemref{spanning}. By
    \lemref{inherent}, $t_i \leq 2t$, for all $i$. As such, the
    crossing number of $T$ is $O\pth{ \MakeBig \sum_i ( t_i + \log n
       /\log \log n ) } = O(t \log n + \log^2 n / \log \log n)$.
\end{proof}

\medskip

When $\PntSet$ is a set of points in $\Re^d$, we will be interested in
the spanning tree having the minimal number of crossings with any
hyperplane. In particular, the above result implies the following.

\begin{corollary}
    Let $\PntSet$ be a set of $n$ points in $\Re^d$, and $t$ be the
    minimum crossing number of any spanning tree of $\PntSet$.  Then,
    one can compute, in polynomial time, a spanning tree $T$ of
    $\PntSet$ with a crossing number $\Delta = O(t \log n + \log^2 n
    /\log \log n )$.  Specifically, any hyperplane in $\Re^d$ crosses
    at most $\Delta$ edges (i.e., segments) of $T$.

    \corlab{low:crossing} 
\end{corollary}

\section{Spanning tree in \TPDF{$\Re^d$}{Rd} with low crossing number}
\seclab{planar}

Let $\PntSet$ be a set of $n$ points in the plane in general position
(i.e., no three points are colinear). Let $\LinesX{\PntSet}$ denote
the set of all partitions of $\PntSet$ into two non-empty sets, by a
line that does not contain any point of $\PntSet$. For each such
partition, we select a representative line that realizes this
partition. We slightly abuse notations as refers to $\LinesX{\PntSet}$
as a set of these lines.

We are interested in the question of finding a spanning tree $T$ of
$\PntSet$ such that each line of $\LinesX{\PntSet}$ crosses at most
$O(\sqrt{n})$ edges of $T$.


\begin{defn}
    For a set of lines $\LineSet$ in the plane, the crossing distance
    between two points is the number of lines of $\LineSet$ crossed by
    the segment formed by these two points. Formally, for any two
    points $\pnt, \pntA \in \Re^2$, the \emphi{crossing distance}
    between them is $\dCr{\LineSet}{\pnt, \pntA} = x+y/2$, where $x$
    is the number of lines of $\LineSet$ having $\pnt$ and $\pntA$ on
    opposite sides, and $y$ is the number of lines that contain either
    $\pnt$ or $\pntA$. It is easy to verify that
    $\dCr{\LineSet}{\cdot}$ complies with the triangle inequality (as
    such its a pseudo-metric).
    
    The \emphi{crossing disk} of radius $r$ centered at a point
    $\pnt$, is the set of all vertices of the arrangement 
    $\ArrX{\LineSet}$ in crossing distance at most $r$ from $\pnt$. We
    denote this ``disk'' by $\diskCr{\LineSet}{\pnt,r}$.
\end{defn}

We need the following lemma due to Welzl \cite{w-stlcn-92}.
\begin{lemma}[\cite{w-stlcn-92}]
    Let $r\geq 0$ be a parameter, $\LineSet$ be a set of lines (of
    size at least $2r$) in the plane, and let $\pnt$ be a point in the
    plane not contained in any line of $\LineSet$. Then
    $\cardin{\diskCr{\LineSet}{\pnt, r}} \geq \binom{r+1}{2}$.

    \lemlab{crossing:disk}
\end{lemma}

Here is the \LP \EqrefQ{l:p:general} specialized for this planar case,
and its dual \LP.

\medskip

\noindent\fbox{\begin{minipage}{0.45\linewidth}
\begin{align*}
    \gamma'(\PntSet,t) &= \max   \sum_{\pnt\pntA \in \pairsX{\PntSet}}
    x_{\pnt \pntA} & \\
    s.t. \quad\quad & \sum_{\substack{\pnt \pntA \in \pairsX{\PntSet},
          \\
          \pnt\pntA
       \cap \ell \ne \emptyset}}  x_{\pnt\pntA} \leq t & \forall \ell \in
    \LinesX{\PntSet} \\
    & \sum_{\pntA \in \PntSet, \pntA \ne \pnt} x_{\pnt\pntA} \geq 1 &
    \forall \pnt \in \PntSet\\
    & x_{\pnt\pntA} \geq 0 & \forall \pnt\pntA \in \pairsX{\PntSet}.
\end{align*}
\end{minipage}}~~
\fbox{\begin{minipage}{0.45\linewidth}
\begin{align*}
    \alpha'(\PntSet,t) & = \min \;\;\; t \sum_{\ell \in \LinesX{\PntSet}}
    z_\ell - \sum_{\pnt \in \PntSet} z_\pnt      & \\
    s.t. \quad\quad & \sum_{\substack{\ell \in \LinesX{\PntSet},
          \\
          \ell \cap \pnt\pntA \ne \emptyset}} z_{\ell} - z_\pnt -
    z_\pntA \geq 1 
    \\[-0.75cm]
    & 
    \hspace{4cm}
    \forall \pnt \pntA \in
    \pairsX{\PntSet} 
    \hspace{-4cm}
    \\[0.2cm]
    & z_{\ell} \geq 0 & \hspace{-4cm} \forall \lineA \in \LinesX{\PntSet}\\
    & z_{\pnt} \geq 0 & \forall \pnt \in \PntSet.
\end{align*}
\end{minipage}}

\medskip

We will next show that $\gamma'(\PntSet, \sqrt{n} )$ is feasible. This
would imply that one can find spanning graph of $\PntSet$ with
crossing number of $O(\sqrt{n})$ that uses a constant fraction of the
vertices (i.e., \lemref{spanning}).

\begin{lemma}
    The \LP $\gamma'(\PntSet, t)$ is feasible for $t = \sqrt{n}$.

    \lemlab{feasible}
\end{lemma}
\begin{proof}
    Consider the dual \LP above and observe that it is always feasible
    (for example by setting $z_\lineA =1$ for all $\lineA \in
    \LinesX{\PntSet}$ and $z_\pnt = 0$ for all $\pnt \in \PntSet$).
    Thus, if we show that $\alpha'(P,t)$ is bounded from below (and
    thus is finite), then the strong duality theorem would imply that
    $\gamma'(\PntSet,t)$ is feasible and equal to $\alpha'(P,t)$.

    So consider a solution to this dual \LP, where all the values are
    rational numbers. Let $U>1$ be the smallest integer such that if
    we scale all the values in the given \LP solution by $U$ then they are
    integers. In particular, let $y_\lineA =  U z_\lineA$, for all
    $\lineA \in \LinesX{\PntSet}$, and $y_\pnt = U z_\pnt$, for all
    $\pnt \in \PntSet$.

    Let $\LineSet$ be a set of lines, where we pick $y_\lineA$ copies
    of $\lineA$ into this set, for all $\lineA \in
    \LinesX{\PntSet}$. Formally, $\psi$ copies of the same line
    $\lineA$ (put into $\LineSet$) will be a collection of $\psi$,
    almost identical, copies of the line $\lineA$ slightly perturbed
    so that these $\psi$ lines are in general position.  Thus,
    $\LineSet$ is a set of $N = U \sum_{\lineA \in \LinesX{\PntSet}}
    z_\lineA$ lines in general position. Furthermore, the inequalities
    in the \LP implies that, for any segment $\pnt \pntA \in
    \pairsX{\PntSet}$, we have that $\pnt \pntA$ crosses
    \[
    \dCr{L}{\pnt, \pntA} =
    \sum_{\substack{\ell \in \LinesX{\PntSet},
          \\
          \ell \cap \pnt\pntA \ne \emptyset}} y_{\ell} 
    = 
    U \sum_{\substack{\ell \in \LinesX{\PntSet},
          \\
          \ell \cap \pnt\pntA \ne \emptyset}} z_{\ell} 
    \geq U \; \pth{1 + z_\pnt +
       z_\pntA} = U + y_\pnt + y_\pntA
    \]
    lines of $\LineSet$. 

    Observe that $\diskCr{\LineSet}{\pnt, y_\pnt} \;\cap\;
    \diskCr{\LineSet}{\pntA, y_\pntA} = \emptyset$ for any pair $\pnt
    \pntA \in \pairsX{\PntSet}$. Otherwise, there would be a point
    $\pntB$ in the plane such that $\dCr{L}{\pnt, \pntB} \leq y_\pnt$
    and $\dCr{L}{\pntA, \pntB} \leq y_\pntA$. But the triangle
    inequality would imply that $\dCr{L}{\pnt, \pntA} \leq y_\pnt +
    y_\pntA$, which contradicts the above.
      
    By \lemref{crossing:disk}, for $\pnt \in \PntSet$, the disk
    $\diskCr{\LineSet}{\pnt, r}$ contains at least
    $\binom{y_\pnt+1}{2}$ distinct vertices of $\ArrX{\LineSet}$. On
    the other hand, the total number of vertices in the arrangement
    $\ArrX{\LineSet}$ is $\binom{N}{2}$. We conclude that
    \[
    \frac{U^2}{2} \sum_{\pnt \in \PntSet} z_\pnt^2 \leq \sum_{\pnt \in
       \PntSet} \binom{y_\pnt+1}{2} \leq \binom{N}{2} \leq \frac{U^2}{2}
    \pth{\sum_{\lineA \in\LineSet} z_\lineA}^2.  \;\;\; \implies
    \;\;\; \sum_{\pnt \in \PntSet} z_\pnt^2 \leq  \pth{\sum_{\lineA
          \in\LineSet} z_\lineA}^2.
    \]
    Now, by the Cauchy-Schwarz inequality and the above, we have that
    \[
    \sum_{\pnt \in \PntSet} z_\pnt \leq \sqrt{n}\sqrt{ \sum_{\pnt \in
          \PntSet} z_\pnt^2}
    \leq
    \sqrt{n} \sum_{\lineA \in\LineSet} z_\lineA.    
    \]
    As $t=\sqrt{n}$ this implies that $\alpha'(n,t) = \sqrt{n}
    \sum_{\lineA \in\LineSet} z_\lineA - \sum_{\pnt \in \PntSet}
    z_\pnt \geq 0$. The claim now follows.
\end{proof}


\begin{remark}
    The proof of \lemref{feasible} works also in higher dimensions,
    where we consider points and hyperplanes in $\Re^d$. There, one
    has to use \Holder's inequality instead of the Cauchy-Schwarz
    inequality. Then, the \LP is feasible for $t = O\pth{
       n^{1-1/d}}$.
    \remlab{easy}
\end{remark}

\begin{theorem}
    Given a set $\PntSet$ of $n$ points in $\Re^d$, one can compute
    (in polynomial time) a spanning tree $T$ of $\PntSet$ with crossing
    number at most $O\pth{ n^{1-1/d}}$; that is, any hyperplane in
    $\Re^d$ crosses at most $O\pth{ n^{1-1/d}}$ edges of $T$.
\end{theorem}
\begin{proof}
    By \lemref{feasible} and \remref{easy}, $\gamma'(\PntSet, t )$ is
    feasible, for $t =O\pth{ n^{1-1/d}}$. As such, by
    \lemref{spanning}, we can compute a set of edges $F$ that engages
    a constant fraction of the points of $\PntSet$, and it has
    crossing number $O( t + \log n / \log \log n ) = O(t)$.  Using the
    algorithm of \thmref{low:crossing} generates a spanning tree with
    crossing number $C(n) = O\pth{n^{1-1/d}} + C((19/20)n )$. Namely,
    the resulting spanning tree has crossing number
    $O\pth{n^{1-1/d}}$.
\end{proof}


\begin{remark}[Connection to separating/hitting and packing \LP{}s.]
    It is interesting to consider the \LP that just
    tries to separate all points of $\PntSet$ from each other. It
    looks similar to our dual \LP while being simpler.
    
    \smallskip

    \centerline{\fbox{\begin{minipage}{0.45\linewidth}
              \vspace{-0.35cm}
              \begin{align*}
                  \mathrm{sep}(\PntSet) & = \min \;\;\; \sum_{\ell \in
                     \LinesX{\PntSet}}
                  z_\ell      & \\
                  s.t. \quad\quad & \sum_{\substack{\ell \in \LinesX{\PntSet},
                        \\
                        \ell \cap \pnt\pntA \ne \emptyset}} z_{\ell} \geq 1
                  &  \forall \pnt \pntA \in \pairsX{\PntSet}
                  \\[0.2cm]
                  & z_{\ell} \geq 0 & \hspace{-4cm} \forall \lineA \in \LinesX{\PntSet}\\
                  & z_{\pnt} \geq 0 & \forall \pnt \in \PntSet.
              \end{align*}
          \end{minipage}}}

    \smallskip

    Now, a similar scaling argument to the one used in the proof of
    \lemref{feasible} implies that $\sum_{\lineA \in \LinesX{\PntSet}}
    z_\lineA \geq \sqrt{n}/2$. Namely, any fractional set of lines
    separating $n$ points in the plane is of size $\Omega(\sqrt{n})$.
    Naturally, this argument works also in higher dimensions, where a
    fractional set of hyperplanes of size $\Omega(n^{1-1/d})$ is
    required to separate a set of  $n$ points in $\Re^d$.

    Observe, that since the \LP $\alpha'(\gamma, t)$ is more
    restrictive than this \LP, we conclude that for any feasible
    solution to $\alpha'(\gamma, t)$ it holds that $\sum_{\lineA \in
       \LinesX{\PntSet}} z_\lineA \geq \sqrt{n}/2$.

    The dual to the above \LP is the packing \LP that tries to pick as
    many fractional edges as possible, while no line crosses edges
    with total value exceeding $1$. While this is similar to our
    primal \LP $\gamma'(\PntSet,t)$, it is not clear how to round it,
    since we do not have the guarantee that every point has sufficient
    number of edges attached to it in the fractional solution.
\end{remark}



\subsection{A Deterministic algorithm for the planar case}
\seclab{deterministic}

\medskip

\SaveIndent
\noindent
\begin{minipage}{0.49\linewidth}
    \RestoreIndent 

    Interestingly, at least in the planar case, one can
    do the rounding deterministically.

    \begin{lemma}
        Let $\PntSet$ be a set of $n$ points in the plane and a parameter
        $t$, such that $\gamma'(\PntSet,t)$ is feasible.  Then, one can
        compute, in polynomial deterministic time, a set of edges $F$,
        such that (i) the crossing number of $F$ is $\leq 12 t$, and (ii)
        the number of connected components in $(\PntSet,F)$ is $\leq
        (3/4)n$.

        \lemlab{planar}
    \end{lemma}       
\end{minipage}~
\begin{minipage}{0.47\linewidth}
    \vspace{-0.2cm}
    \fbox{
       \begin{minipage}{0.97\linewidth}           
           \vspace{-0.4cm}
           \begin{align}
               \eqlab{second:l:p}
               \min& \sum_{\pnt \pntA \in
                  \pairsX{\PntSet} } \dist{ \pnt - \pntA} x_{\pnt
                  \pnt} &\\
     s.t. \quad & \sum_{\substack{\pnt \pntA \in \pairsX{\PntSet},
          \\
          \pnt\pntA
       \cap \ell \ne \emptyset}}  x_{\pnt\pntA} \leq t & \forall \ell \in
 \LinesX{\PntSet} \nonumber \\
  & \sum_{\pntA \in \PntSet, \pntA \ne \pnt}
               x_{\pnt\pntA} \geq 1 &
               \forall \pnt \in \PntSet
               \tag{*}\\
               & x_{\pnt\pntA} \geq 0 &
               \!\!\!\!\!\!\!\!\!\!\!\!\!\!\!\!\!\!\!
               \forall \pnt\pntA \in
               \pairsX{\PntSet}.\nonumber
           \end{align}
       \end{minipage} 
    }

    \vspace{-0.2cm}

    \caption{The modified \LP}
    \figlab{modified:l:p}
\end{minipage} 

\medskip

\begin{proof}
    Instead of computing $\rho =
    \gamma'(\PntSet, t)$ we slightly modify the \LP so that it
    finds the ``shortest'' such solution.  The resulting modified
    \LP is depicted in 
    \figref{modified:l:p}.

    
    \parpic[r]{\includegraphics{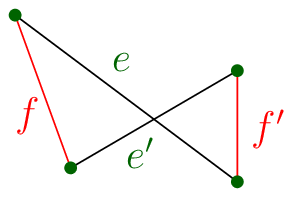}}

    Let $H$ be the set of all the edges $\pnt \pntA$ in the solution
    to this \LP such that $x_{\pnt\pntA} > 0$. We claim that this set
    of edges is planar. Indeed, if two such segments $e$ and $e'$
    intersect, then consider two opposing edges $f$ and $f'$ of the
    quadrant formed by the convex hull of the endpoints of $e$ and
    $e'$, see figure on the right.

    We have that $\dist{f} + \dist{f'} < \dist{e} + \dist{e'}$, which
    implies that, for $\delta >0$ sufficiently small, the solution
    $x_e = x_e -\delta$, $x_{e'} = x_{e'} -\delta$, $x_{f} = x_{f}
    +\delta$, $x_{f'} = x_{f'} +\delta$ is feasible, as the total
    value of the edges attached to a vertex does not change, and the
    crossing number of any line does not increase by this change, as
    can be easily verified. But this implies that there is a feasible
    solution with a better target value (specifically, the target
    value goes down by $\delta \cdot (\dist{e} + \dist{e'} - \dist{f} -
    \dist{f'})$). A contradiction.

    Thus $G = (\PntSet, H)$ is a planar graph where each point of
    $\PntSet$ has at least one edge attached to it. Furthermore, the
    average degree in a planar graph is at most $6$, which implies
    that at least half of the points of $\PntSet$ have degree at most
    $12$ in $G$. But each such point $\pnt$, must have an edge $\pnt
    \pntA$ attached to it, such that $x_{\pnt \pntA} \geq 1/12$,
    because of (*) in the \LP \EqrefQ{second:l:p}.
    
    Thus, scale the \LP solution by a factor of $12$ and pick all the
    edges $\pnt \pntA$ with $12 x_{\pnt \pntA} \geq 1$ into the set
    $F$. We get a set that is adjacent to at least half of the points
    of $\PntSet$, and its crossing number is at most $12 t$.
\end{proof}    

\medskip

The planarity argument in the above proof of \lemref{planar} is
similar to the one used by Fekete \etal \cite{flm-msnmt-08} -- they
use it to argue that there is one heavy edge, while we use it to argue
that there are many heavy edges.

\begin{theorem}
    Let $\PntSet$ be a set of $n$ points in the plane.  One can
    compute, in deterministic polynomial time, a spanning tree $T$ of
    $\PntSet$ with a crossing number $O(\min( t \log n, \sqrt{n}) )$,
    where $t$ is the minimum crossing number of any spanning tree of
    $\PntSet$.

    \thmlab{low:crossing:2}
\end{theorem}

\section{Conclusions}
\seclab{conclusions}

We presented an approximation algorithm for computing a spanning tree
with low crossing number. The new algorithm relies on a natural \LP
relaxation of the problem and a straightforward rounding scheme.

Interestingly, our approach enables us to provide a direct proof to
the existence of such spanning trees in $\Re^d$. This is, as far as we
know, the first algorithm for this problem that avoids using the
reweighting technique.  Intuitively, our algorithm (together with
previous results \cite{l-upsda-01}) suggests that reweighting in
geometric settings can sometimes be replaced by \LP rounding. This is
a significant feature, as \LP{}s are considerably more general and
flexible tool than reweighting. For example, using our algorithm, we
can add other constraints to the \LP; e.g., we can insist that some
certain cuts would have significantly lower crossing number than some
other cuts. In particular, it is not clear how one can incorporate
such considerations into a reweighting algorithm computing spanning
trees with low crossing number.

One interesting open problem, is to compute spanning trees with
\emph{relative} crossing number using the new \LP approach. Here,
given a point set $\PntSet$ in $\Re^3$, one would like to compute a
spanning tree $T$ such that if a halfspace $h^+$ contains $k$ points
of $\PntSet$ then it boundary plane $h$ crosses (say) $O( (k \log
n)^{2/3})$ edges of $T$. Such a result is known in the plane
\cite{ahs-ahrcr-07}, but the problem is open in higher dimensions.

\section*{Acknowledgments}

The author would like to thank Chandra Chekuri, S{\'a}\si{ndor}
Fekete, Jirka \Matousek, and Emo Welzl for helpful discussions on the
problems studied in this paper.

 
\bibliographystyle{alpha} 
\bibliography{shortcuts,geometry}

\end{document}